\documentclass[aps,superscriptaddress, nofootinbib, showpacs,12pt]{revtex4}
\usepackage{amsmath,amssymb,amsfonts, bm, natbib}
\usepackage{dsfont}
\usepackage{epsfig}
\usepackage{graphicx}
\usepackage{slashed}
\usepackage{color}

\usepackage{caption}
\usepackage{subcaption}
\usepackage{slashed}

\usepackage{commath}
\usepackage{calc}

\newcommand{\be}{\begin{equation}}
\newcommand{\ee}{\end{equation}}

\newcommand{\bea}{\begin{eqnarray}}
\newcommand{\eea}{\end{eqnarray}}

\newcommand{\bfig}{\begin{figure}}
\newcommand{\efig}{\end{figure}}

\makeatletter
\newcommand*{\rom}[1]{\expandafter\@slowromancap\romannumeral #1@}
\makeatother

\begin{document}

\title{Low scale left-right-right-left symmetry}
\author{Gauhar Abbas}
\email{Gauhar.Abbas@ific.uv.es}
\affiliation{IFIC, Universitat de Val\`encia -- CSIC, Apt. Correus 22085, 
E-46071 Val\`encia, Spain}

\begin{abstract}
We propose an effective left-right-right-left model with a parity breaking scale around a few TeV.   One of the  main achievements of the model is that the mirror fermions as well as  the mirror gauge sector simultaneously could be at TeV scale.  It is shown that the most dangerous quadratic divergence of the SM Higgs boson  involving the top quark in the loop is naturally suppressed, and begins at three loop.  The model postpones the fine-tuning of the mass of the SM Higgs boson up to a sufficiently high scale.  The model explains the smallness of the neutrino masses whether they are Dirac or Majorana. Furthermore, the strong $CP$ phase is zero in this model.
\end{abstract}


\maketitle

Left-right-right-left (LRRL) models are an alternative and elegant way of restoring parity at a high scale\cite{Abbas:2016xgj}.  In these models, the standard model (SM) left- and right-handed fermions are kept in the fundamental representation of the gauge groups $SU(2)_L$  and $SU(2)_R$, respectively.  This is similar to left-right symmetric (LRS) models\cite{Pati:1974yy,Mohapatra:1974hk,Senjanovic:1975rk,Senjanovic:1978ev}.  However, the coupling constants of the gauge groups $SU(2)_L$  and $SU(2)_R$ are independent.  Now, the question is if parity can be restored.  The  simplest possibility is to assume that there are gauge symmetries $SU(2)^{\prime}_R$ and $SU(2)^{\prime}_L$ which are  parity or mirror counterparts of the gauge groups $SU(2)_L$ and $SU(2)_R$, respectively.  This is also the simplest way to  introduce new fermions to the SM in this scenario.  These are  unique features of LRRL models.  Furthermore, the scalar sector of LRRL models is elegant and optimum. 

On the phenomenological side, LRRL models have a good motivation from a recently observed excess by the ATLAS and CMS collaborations\cite{Aad:2015owa,Aad:2014xka,Aad:2015ufa,Khachatryan:2014hpa,Khachatryan:2014gha,Khachatryan:2014dka,CMS:2015gla,Khachatryan:2015sja,Aad:2014aqa}. 
This is established that this excess  can be explained with different coupling constants for $SU(2)_L $ and $ SU(2)_R $\cite{Dobrescu:2015qna,Dobrescu:2015yba,Deppisch:2014qpa,Deppisch:2014zta,Heikinheimo:2014tba,Aguilar-Saavedra:2014ola,Fowlie:2014mza,Krauss:2015nba,Gluza:2015goa,Dev:2015pga}\footnote{With new data, all these excesses have disappeared.}.  LRRL models also provide an interesting perspective from the pure theoretical point of view.  We note that the symmetry   $SU(3)_c \otimes SU(2)_L \otimes SU(2)_R \otimes SU(2)^{\prime}_R \otimes SU(2)^{\prime}_L \otimes U(1)_{Y} $ of LRRL models cannot be embedded in $SU(5)$ or $SO(10)$ type GUT models. LRRL models  might present an interesting possibility  for a new and a low scale unification scenario.  For example, the nearest unification could come from a $SU(4)_1 \otimes SU(4)_2$ type of model where $SU(2)_L$ and $SU(2)^{\prime}_R$ can be embedded in $SU(4)_1$ whereas $SU(2)_R$ and $SU(2)^{\prime}_L$  can live inside $SU(4)_2$. 

However, models based on mirror fermions and mirror symmetries come with a great disadvantage\cite{Foot:1991bp,Silagadze:1995tr,Foot:1995pa,Berezhiani:1995yi,Gu:2012in,Chakdar:2013tca}. Parity invariance dictates that  the Yukawa couplings of the mirror fermions should be identical to that of the SM ones.  The LHC has not found these mirror fermions around TeV scale yet.  Hence, for keeping the masses of mirror fermions at TeV scale, parity breaking scale should be very high ($ 10^{8}$ GeV or so)\cite{Foot:1991bp,Silagadze:1995tr,Foot:1995pa,Berezhiani:1995yi,Gu:2012in,Chakdar:2013tca}.  This raises the scale of the mirror gauge sector to, for example, $10^8$ GeV.  Thus, the new mirror gauge sector of these models is out of the reach of the LHC, and it is practically impossible to produce mirror gauge bosons  with present day technologies.

In this paper, we investigate whether it is possible to have mirror gauge sector as well as  the mirror fermions around TeV scale within the framework of the LRRL symmetry\cite{Abbas:2016xgj}.  We propose a new type of LRRL model  which provides a low scale parity breaking  resulting in a low scale mirror gauge sector as well as low scale mirror fermions.  This is one of the main achievements of this proposed work which is near impossible in other models having mirror fermions and mirror gauge symmetries\cite{Foot:1991bp,Silagadze:1995tr,Foot:1995pa,Berezhiani:1995yi,Gu:2012in,Chakdar:2013tca}.  Furthermore, we shall see that the dangerous quadratic divergences of the SM Higgs mass involving fermions loops are suppressed, and begin at the three loop level.   The model can stabilize the mass of the SM Higgs up to a sufficiently high scale.

The fermionic and gauge fields under  parity transform as the following in LRRL models:
\begin{eqnarray}
&&\psi_L \longleftrightarrow \psi^{\prime}_R,~ \psi_R \longleftrightarrow \psi^{\prime}_L, ~
\mathcal{W}_L \longleftrightarrow  \mathcal{W}_R^\prime,~
 \mathcal{W}_R \longleftrightarrow  \mathcal{W}_L^\prime,~
  \mathcal{B}_\mu \longleftrightarrow  \mathcal{B}_\mu,~\mathcal{G}_{\mu \nu} \longleftrightarrow  \mathcal{G}_{\mu \nu},
\end{eqnarray}
where $\psi_L$, $\psi_R$ are doublets of the  gauge groups $SU(2)_L$ and  $SU(2)_R$, respectively.  The doublets $\psi^{\prime}_L$, $\psi^{\prime}_R$ correspond to the gauge groups $SU(2)^{\prime}_L$ and  $SU(2)^{\prime}_R$, respectively. $ \mathcal{W}_L $ and $ \mathcal{W}_R $ are the gauge fields corresponding to $SU(2)_L$ and $SU(2)_R$, respectively.  $ \mathcal{W^\prime}_R $ and $ \mathcal{W^\prime}_L$ are gauge fields of the symmetries $SU(2)^{\prime}_R$ and $SU(2)^{\prime}_L$, respectively. $\mathcal{B}_\mu $ is the gauge field corresponding to the gauge symmetry group $U(1)_Y$.  $\mathcal{G}_{\mu \nu}$ is gluon field strength tensor representing the $SU(3)_c$ color symmetry.

The fermions of the model have the following transformations under the symmetry $SU(3)_c \otimes SU(2)_L \otimes SU(2)_R \otimes SU(2)^{\prime}_R \otimes SU(2)^{\prime}_L \otimes U(1)_{Y} $:
\begin{eqnarray}
&&Q_L :(3,2,1,1,1,\frac{1}{3} ),~Q_R :(3,1,2,1,1,\frac{1}{3} ),~
Q_R^\prime:(3,1,1,2,1,\frac{1}{3} ),~Q_L^\prime :(3,1,1,1,2,\frac{1}{3} ), \\ \nonumber
&&L_L :(1,2,1,1,1,-1),~L_R :(1,1,2,1,1,-1),~
L_R^\prime:(1,1,1,2,1,-1 ),~L_L^\prime :(1,1,1,1,2,-1 ),
 \end{eqnarray} 
where $Q$ and $L$ denote the quarks and leptonic doublets.  For more details, see Ref. \cite{Abbas:2016xgj}.

We introduce four Higgs doublet and two singlet real scalar fields for the spontaneous symmetry breaking (SSB) which transform in the following way under $SU(3)_c \otimes SU(2)_L \otimes SU(2)_R \otimes SU(2)^{\prime}_R \otimes SU(2)^{\prime}_L \otimes U(1)_{Y} $:

\begin{eqnarray}
&&\varphi_L :(1,2,1,1,1,1),~\varphi_R :(1,1,2,1,1,1),~
\varphi_R^\prime:(1,1,1,2,1,1), \\ \nonumber
&&\varphi_L^\prime :(1,1,1,1,2,1),~
 \chi:(1,1,1,1,1,0),~ \chi^\prime:(1,1,1,1,1,0).
 \end{eqnarray} 
The scalar fields under parity behave as follows:
\begin{eqnarray}
&&\varphi_L \longleftrightarrow \varphi^{\prime}_R,~
\varphi_R \longleftrightarrow \varphi^{\prime}_L,~
\chi \longleftrightarrow \chi^{\prime}.
\end{eqnarray}

Now, the SSB occurs in the following pattern:  The vacuum expectation value (VEV) of the scalar fields $\varphi_L^\prime$ breaks the whole symmetry $SU(2)_L \otimes SU(2)_R \otimes SU(2)^{\prime}_R \otimes SU(2)^{\prime}_L \otimes U(1)_{Y}$   to $SU(2)_L \otimes SU(2)_R \otimes SU(2)^{\prime}_R \otimes  U(1)_{Y^{\prime}}$. 
After this, we break  $SU(2)_L \otimes SU(2)_R \otimes SU(2)^{\prime}_R \otimes  U(1)_{Y^{\prime}}$  to $SU(2)_L \otimes SU(2)_R \otimes  U(1)_{Y^{\prime \prime}}$ using the VEV of the scalar field $\varphi_R^\prime$. 
The $SU(2)_L \otimes SU(2)_R \otimes  U(1)_{Y^{\prime \prime}}$ is broken down to the SM gauge group $SU(2)_L \otimes  U(1)_{Y^{\prime \prime \prime}}$ by the VEV of the scalar field  $\varphi_R$
Finally, the VEV of  the scalar field $\varphi_L$ breaks the SM gauge symmetry   $SU(2)_L \otimes  U(1)_{Y^{\prime \prime \prime}}$  to the $U(1)_{EM}$.

The Yukawa Lagrangian does not exist  since there is no bidoublet in this model. Now, the only way to give masses to fermions  is to use nonrenormalizable operators which makes this model an effective theory.  For this purpose, we observe that all nonrenormalizable operators are ``equal".  For example, due to given scalar fields of the model, we could use a dimension-$5, 6,7 \cdots$ or any  operator  for this purpose.

However, our aim is to have mirror fermions and mirror gauge sector at the same scale which could be around a few TeV.  For this purpose, we demand that fermionic fields $\psi_R$, $\psi_L^\prime$ and scalar singlets $\chi$, $\chi^\prime$ transform under two discrete symmetries, $\mathcal{Z}_2$ and  $\mathcal{Z}_2^\prime$ as given in Table \ref{tab1}.  All other fields are even under $\mathcal{Z}_2$ and  $\mathcal{Z}_2^\prime$.

\begin{table}[h]
\begin{center}
\begin{tabular}{|c|c|c|}
  \hline
  Fields             &        $\mathcal{Z}_2$                    & $\mathcal{Z}_2^\prime$        \\
  \hline
  $\psi_R$                 &   +  &     -                                    \\
  $\chi$                        & +  &      -                                                   \\
  $ \psi_L^\prime$     & -  &   +                                              \\
  $\chi^\prime$           & - &      +            \\
  \hline
     \end{tabular}
\end{center}
\caption{The charges of fermionic and singlet scalar fields under $\mathcal{Z}_2$ and  $\mathcal{Z}_2^\prime$  symmetries.}
 \label{tab1}
\end{table}

Now, the mass term for charged fermions appears at dimension-6.  Thus, the mass term for quarks is given by

\be
\label{mass1}
{\mathcal{L}}^{Q}_{mass} = \dfrac{1}{\Lambda^2} \left[ \bar{Q_L} \left( \Gamma_1   \varphi_L \varphi_R^\dagger \chi + \Gamma_2 \tilde{\varphi}_L  \tilde{\varphi}_R^\dagger \chi  \right) Q_R + \bar{Q^{\prime}_{R}}  \left( \Gamma_1^{\prime }  \varphi_R^\prime \varphi_L^{\prime \dagger} \chi^\prime  + \Gamma_2^{\prime} \tilde{\varphi}_R^\prime  \tilde{\varphi}_L^{\prime \dagger} \chi^\prime  \right) Q^{\prime}_{L} \right] + H.c.,
\ee
where $\Gamma_i = \Gamma_i^{\prime}$ due to parity and $ \tilde{\varphi} = i \sigma_2 \varphi^*$ is charge conjugated Higgs field.  A similar Lagrangian can be written for leptons. Parity is spontaneously broken when scalar fields acquire vacuum expectation values (VEV) such that $ \langle \chi^\prime \rangle >> \langle \varphi_L^\prime \rangle \geq \langle \varphi_R \rangle \geq \langle \varphi_R^\prime \rangle  >>    \langle \varphi_L \rangle$ and $ \langle \chi^\prime \rangle >> \langle \chi \rangle $.

Now,  let us assume that parity breaking scale is around a few TeV.  This means that the gauge bosons corresponding to the gauge groups $SU(2)_R, SU(2)_R^\prime$  and $SU(2)_L^\prime $ should be around a few TeV.  Since, the Yukawa couplings of the mirror fermions are identical to those of the SM ones, we would naively expect that mirror fermions could be very light and already ruled out by  experiments.  However, we observe that due to the VEV pattern described, the VEV of the singlet $\langle \chi^\prime \rangle$ could be large so that mirror fermions, in spite of a TeV scale parity breaking, could be sufficiently heavy to search at the LHC.   Thus, we observe that the mirror fermions and mirror gauge sector both could be at TeV scale in this model.  This is obtained in a natural way, and is one of the  main achievements of this work.  The LHC has searched for these quarks, and has excluded them up to $690$ GeV.  However, these searches are model dependent\cite{Aad:2015tba}.

We note that the models with mirror fermions discussed in the literature are either based on $SU(2) \otimes U(1) \otimes SU(2)^\prime \otimes U(1)^\prime$ or $SU(2)_L \otimes SU(2)_R \otimes U(1)$ symmetries\cite{Foot:1991bp,Silagadze:1995tr,Foot:1995pa,Berezhiani:1995yi,Gu:2012in,Chakdar:2013tca}.  These models have a well-defined Yukawa Lagrangian.  Any attempt to raise the mass scale of the mirror fermions using singlet scalar fields will kill the Yukawa Lagrangian making these models artificial and unnatural. Furthermore, these models do not yield any explanation for the smallness of neutrino masses.

The Majorana mass term for neutrinos can be written at dimension-5,
\be
\label{mass2}
{\mathcal{L}}^{\nu}_{Majorana} = \dfrac{1}{\Lambda} \left[ \bar{L_{L}^c}  c_1   \tilde{\varphi}_{L}^* \tilde{\varphi}_L^\dagger  L_L + \bar{L_{R}^c}^{\prime}  c_1^\prime   \tilde{\varphi}_R^{* \prime} \tilde{\varphi}_R^{\prime \dagger}  L_R^\prime +  \bar{L_{R}^c}  c_2   \tilde{\varphi}_{R}^* \tilde{\varphi}_R^\dagger  L_R + \bar{L_{L}^c}^{\prime}  c_2^\prime   \tilde{\varphi}_L^{* \prime} \tilde{\varphi}_L^{\prime \dagger}  L_L^\prime \right] + H.c.,
\ee
where $c_i = c_i^{\prime}$ due to parity.  We observe in Eqs. (\ref{mass1}) and (\ref{mass2}) that masses of the neutrinos are suppressed by the scale $\Lambda$.  Hence, even if neutrinos are Dirac in nature (which means that nature has chosen couplings $c_i = c_i^{\prime}= 0$ in Eq. (\ref{mass2}) or Lagrangian in Eq. (\ref{mass2}) is forbidden by some symmetry ), the model can provide an explanation for their small masses.

We can also write the Lagrangian which allows the mirror quarks to decay into the SM ones.  This is given by the following dimension-5 and dimension-7 operators:

\be
\label{mass3}
{\mathcal{L}} = \dfrac{\rho}{\Lambda} \bar{Q_L}    \varphi_L \varphi_R^{\prime \dagger}  Q_R^\prime + \dfrac{\sigma}{\Lambda^3} \bar{Q_L}^\prime    \varphi_L^\prime  \chi \chi^\prime \varphi_R^{ \dagger}  Q_R  + H.c.,
\ee
where $\rho$ and $\sigma$ are dimensionless couplings.  We can write a similar Lagrangian for leptons.

The masses and mixings of gauge bosons are obtained from  the following Lagrangian:
\begin{eqnarray}
{\cal L}_{gauge-scalar}& =& \left({\cal D}_{\mu,L}\varphi_L \right)^\dagger\left({\cal D}^{\mu}_L \varphi_L \right)+ \left( {\cal D}_{\mu,R}^\prime \varphi_R^\prime \right)^\dagger \left({\cal D}^{\mu \prime}_R \varphi_R^\prime \right) \\ \nonumber
&+&  \left({\cal D}_{\mu,R}\varphi_R \right)^\dagger\left({\cal D}^{\mu}_R \varphi_R \right) + \left( {\cal D}_{\mu,L}^\prime \varphi_L^\prime \right)^\dagger \left({\cal D}^{\mu \prime}_L \varphi_L^\prime \right),
\label{ktl}
\end{eqnarray}
where, ${\cal D}_{L,R}~{\rm and}~\cal D_{L,R}^\prime$ are the covariant derivatives given by
\begin{eqnarray}
\mathcal{D}_{\mu,L} (\mathcal{D}_{\mu,R}^\prime) &=& \partial_\mu + i g_1  \dfrac{\tau_a}{2} \mathcal{W}^a_{\mu,L} (\mathcal{W}^{a \prime}_{\mu,R}) +ig^\prime \frac{Y}{2} B_\mu,
\end{eqnarray}
\begin{eqnarray}
\mathcal{D}_{\mu,R} (\mathcal{D}_{\mu,L}^\prime)&=& \partial_\mu + i g_2  \dfrac{\tau_a}{2} \mathcal{W}^a_{\mu,R} (\mathcal{W}^{a \prime}_{\mu,L}) +ig^\prime \frac{Y}{2} B_\mu, 
\end{eqnarray}
 where, $\tau_a$'s are the Pauli matrices. The coupling constant $g_1$ corresponds to gauge groups $SU(2)_L $ and $ SU(2)^{\prime}_R$.  The coupling constant of gauge groups $SU(2)_R $ and $ SU(2)^{\prime}_L$ is $g_2$. The coupling constant of gauge group $U(1)_{Y}$ is $g^\prime$.
 
After the SSB, masses of the  charged gauge bosons are given as
\begin{equation}
M_{W^\pm_{L}}~=~\frac{1}{2}g_1 v_L,~~M_{ W^{\prime \pm}_{R}}~=~\frac{1}{2} g_1 v^{\prime}_R,~~M_{W^\pm_{R}}~=~\frac{1}{2}g_2 v_R,~~M_{ W^{\prime \pm}_{L}}~=~\frac{1}{2} g_2 v^{\prime}_L.
\end{equation} 

The nondiagonal mass matrix for the neutral gauge bosons, in the basis ($W^3_L,~W^{\prime 3}_R,~W^3_R,~W^{\prime 3}_L,~B$),  is given by
\begin{equation}
M^2=\frac{1}{4}{\begin{pmatrix} g_1^2 v_L^2 & 0  & 0 & 0 & -g_1 g^\prime v_L^{2} \\ 
0 & g_1^2  v_R^{\prime 2} & 0& 0 & -g_1 g^\prime  v_R^{\prime 2} \\ 
0 & 0 & g_2^2  v_R^2 & 0 & -g_2 g^\prime  v_R^2 \\ 
0 & 0 & 0 & g_2^2  v_L^{\prime 2}  & -g_2 g^\prime v_L^{\prime 2} \\ 
-g_1 g^\prime v_L^{2}  &   -g_1 g^\prime  v_R^{\prime 2} & -g_2 g^\prime  v_R^2 &  -g_2 g^\prime v_L^{\prime 2} & g^{\prime 2}(v_L^2+  v_R^{\prime 2} +  v_R^2 +  v_R^{\prime 2} ) \end{pmatrix}}.\label{nmm}
\end{equation}

This mass matrix can be diagonalized through an orthogonal transformation $R$ which transforms the weak eigenstates ($W^3_L,~W^{\prime 3}_R,~W^3_R,~W^{\prime 3}_L,~B$) to the physical mass eigenstates ($Z_L,~Z_R^\prime,~ Z_R,~ Z_L^\prime,~\gamma$);
\begin{equation}
{\begin{pmatrix} W^3_L  \\ W^{\prime 3}_R \\W^3_R \\ W^{\prime 3}_L  \\ B \end{pmatrix}}=R {\begin{pmatrix} Z_L\\ Z_R^\prime \\\ Z_R \\ Z_L^\prime  \\ \gamma \end{pmatrix}}.
\end{equation}   
 
The physical masses of neutral gauge bosons are given as 
\begin{eqnarray}
M_{Z_L}^2 & = & \dfrac{1}{2} g_1^2 v_L^2  \dfrac{ \left(2 g_2^2 g^{\prime 2} + g_1^2 (g_2^2 + g^{\prime 2} ) \right) }{\left(g_2^2 g^{\prime 2}  + g_1^2 (g_2^2 + g^{\prime 2}  ) \right)} + \mathcal{O} (\epsilon_1, \epsilon_2, \epsilon_3), ~
M_{Z_R^\prime}^2  =  \dfrac{1}{2}  v_R^{\prime 2} \dfrac{ \left( g_2^2 g^{\prime 2} + g_1^2 (g_2^2 + 2 g^{\prime 2} ) \right)}{ (g_2^2 + 2 g^{\prime 2})} + \mathcal{O} (\epsilon_1, \epsilon_2, \epsilon_3) \\ \nonumber 
M_{Z_R}^2 & = &  v_R^{ 2}  \dfrac{ g_2^4 g^{\prime 4} }{ (g_2^2 + g^{\prime 2}) (g_2^2 g^{\prime 2} + g_1^2 (g_2^2 + g^{\prime 2} )) } + \mathcal{O} (\epsilon_1, \epsilon_2, \epsilon_3),~
M_{Z_L^\prime}^2 =   v_L^{\prime 2}  \dfrac{ g^{\prime 4} }{  (g_2^2 + g^{\prime 2} ) } + \mathcal{O} (\epsilon_1, \epsilon_2, \epsilon_3), 
\end{eqnarray}
where $\epsilon_1 = v_L^2/ v_L^{\prime 2}$, $\epsilon_2 = v_L^2/ v_R^{\prime 2}$ and $\epsilon_3 = v_L^2/ v_R^{ 2}$ . We have shown only leading order terms  assuming that $v_L^\prime, v_R, v_R^\prime >> v_L$.    
The orthogonal transformation matrix $R$ can be parametrized in terms of four mixing angles $\theta_{W_L}$, $\theta_{W_R^\prime}$, $\theta_{W_R}$ and $\theta_{W_L^\prime}$ which are the following:
\begin{eqnarray}
\label{wth}
{\rm cos}^2\theta_{W_L} &= & \left(\frac{M_{W_L}^2}{M_{Z_L}^2}\right)_{\epsilon_{1,2,3}=0}=\frac{\left(g_2^2 g^{\prime 2}  + g_1^2 (g_2^2 + g^{\prime 2}  ) \right)}{ 2 \left(2 g_2^2 g^{\prime 2} + g_1^2 (g_2^2 + g^{\prime 2} ), \right)} \nonumber \\
{\rm cos}^2 \theta_{W_R^\prime} &= &\left(\frac{M_{W_R^\prime}^2}{M_{Z_R^\prime}^2}\right)_{\epsilon_{1,2,3}=0}=\frac{ g_1^2  (g_2^2 + 2 g^{\prime 2}) }{2  \left( g_2^2 g^{\prime 2} + g_1^2 (g_2^2 + 2 g^{\prime 2} ) \right)},  \nonumber \\
{\rm cos}^2 \theta_{W_R} &= &\left(\frac{M_{W_R}^2}{M_{Z_R}^2}\right)_{\epsilon_{1,2,3}=0}=\frac{  (g_2^2 + g^{\prime 2}) (g_2^2 g^{\prime 2} + g_1^2 (g_2^2 + g^{\prime 2} )) }{4 g_2^2 g^{\prime 4}},  \nonumber \\
{\rm cos}^2 \theta_{W_L^\prime} &= &\left(\frac{M_{W_L^\prime}^2}{M_{Z_L^\prime}^2}\right)_{\epsilon_{1,2,3}=0}=\frac{ g_2^2 (g_2^2 + g^{\prime 2}) }{4  g^{\prime 4}}.
\end{eqnarray}

The Lagrangian in Eq.(\ref{mass3}) introduces mixing between the SM and mirror fermions.  We can diagonalize the mass matrices of charged fermions via biunitary transformation  by introducing two mixing angles. The mass eigenstates of the charged fermions are related to  the gauge eigenstates through the following transformation:
\begin{equation}
\label{fth}
{\begin{pmatrix} f^g \\  { f^\prime}^g \end{pmatrix}}_{L,R}~=~{\begin{pmatrix}{\rm cos}\theta& {\rm sin}\theta \\-{\rm sin}\theta & {\rm cos}\theta \end{pmatrix}}_{L,R}{\begin{pmatrix}  f \\  { f^\prime} \end{pmatrix}}_{L,R}
\end{equation}
where, $f_{L,R}$ are the left- and right-handed component of the SM fermions and $f_{L,R}^\prime$ denote the mirror fermions.

Now we discuss the scalar potential of the model.  We write the most general scalar potential of the model as follows:
\begin{eqnarray}
V(\varphi_L , \varphi_R, \varphi_R^\prime, \varphi_L^\prime, \chi, \chi^\prime) &=& -\mu_1^2 \Bigl( \varphi_L^\dagger \varphi_L   + \varphi_R^{\prime \dagger} \varphi_R^\prime  \Bigr) -  \mu_2^2 \Bigl( \varphi_R^\dagger \varphi_R   + \varphi_L^{\prime \dagger} \varphi_L^\prime  \Bigr) - \mu_3^2 \Bigl( \chi^2 + \chi^{\prime 2}  \Bigr) \\ \nonumber
&+&  \lambda_1 \Bigl(   (\varphi_L^\dagger \varphi_L)^2 +   (\varphi_R^{\prime \dagger} \varphi_R^\prime)^2  \Bigr) +  \lambda_2 \Bigl(   (\varphi_R^\dagger \varphi_R)^2 +   (\varphi_L^{\prime \dagger} \varphi_L^\prime)^2  \Bigr) \\ \nonumber
&+& \lambda_3 \Bigl( \varphi_L^\dagger \varphi_L \varphi_R^\dagger \varphi_R + \varphi_R^{\prime \dagger} \varphi_R^\prime  \varphi_L^{\prime \dagger} \varphi_L^\prime    \Bigr)   
+ \lambda_4 \Bigl(  \varphi_L^\dagger \varphi_L   \varphi_L^{\prime \dagger} \varphi_L^\prime +  \varphi_R^\dagger \varphi_R \varphi_R^{\prime \dagger} \varphi_R^\prime   \Bigr) \\ \nonumber
&+& \lambda_5   \varphi_L^\dagger \varphi_L   \varphi_R^{\prime \dagger} \varphi_R^\prime  + \lambda_6   \varphi_R^\dagger \varphi_R \varphi_L^{\prime \dagger} \varphi_L^\prime   + \lambda_7 \Bigl(  \chi^4 + \chi^{\prime 4}  \Bigr) + \lambda_8  \chi^2 \chi^{\prime 2} \\ \nonumber
&+&  \lambda_9 \Bigl(   \varphi_L^\dagger \varphi_L \chi^2 +   \varphi_R^{\prime \dagger} \varphi_R^\prime \chi^{\prime 2}  \Bigr) + \lambda_{10} \Bigl(   \varphi_L^\dagger \varphi_L \chi^{\prime 2} +   \varphi_R^{\prime \dagger} \varphi_R^\prime \chi^{ 2}  \Bigr) \\ \nonumber
&+&  \lambda_{11} \Bigl(   \varphi_R^\dagger \varphi_R \chi^2 +   \varphi_L^{\prime \dagger} \varphi_L^\prime \chi^{\prime 2}  \Bigr) + \lambda_{12} \Bigl(   \varphi_R^\dagger \varphi_R \chi^{\prime 2} +   \varphi_L^{\prime \dagger} \varphi_L^\prime \chi^{ 2}  \Bigr). 
\end{eqnarray}
The VEVs of the Higgs fields are denoted as $ \langle \varphi_L \rangle = v_L/\sqrt{2}, \langle \varphi_L \rangle = v_R/\sqrt{2}, \langle \varphi_R^\prime \rangle = v_R^\prime/\sqrt{2}, \langle \varphi_L^\prime \rangle =  v_L^\prime/\sqrt{2}, \langle \chi \rangle =  \omega/\sqrt{2}, \langle \chi^\prime \rangle =  \omega^\prime/ \sqrt{2}$. We need a solution of the potential such that $ \langle \chi^\prime \rangle >> \langle \varphi_L^\prime \rangle \geq \langle \varphi_R \rangle \geq \langle \varphi_R^\prime \rangle  >>    \langle \varphi_L \rangle$ and $ \langle \chi^\prime \rangle >> \langle \chi \rangle $.  There are six independent vacuum parameters which correspond to
six independent vacuum minimal conditions, i.e.,
\begin{equation}
0=\frac{\partial V}{\partial v_L}=\frac{\partial V}{\partial v_R}=\frac{\partial V}{\partial v_R^\prime}=\frac{\partial V}{\partial v_L^\prime}=\frac{\partial V}{\partial \omega}=\frac{\partial V}{\partial \omega^\prime}.
\end{equation}

The second derivatives of the scalar potential which is the mass squared matrix determine the nature of the minimum.  This is given by
\be
\label{sec_der}
\frac{\partial^2 V}{\partial ( \varphi_i, \chi; \chi^\prime) \partial ( \varphi_j,  \chi; \chi^\prime)  } > 0.
\ee 

In general, one of the eigenvalue of this matrix is always zero.  Hence, we assume that the mass term for the singlet scalar fields in  the scalar potential is zero ($\mu_3=0$).  This implies that the scalar particles corresponding to the singlet fields $\chi$ and $\chi^\prime$ are mass-less\footnote{We can also assume that only one scalar singlet is massless and the other one is massive.  Then, we need to write a mass term for the other singlet in the potential.  This will break parity softly.}. The reason to choose only singlet scalars to be massless is that they could be dark matter candidates. The phenomenological consequences of this assumption are far reaching and will be discussed in the later course of the paper.

Furthermore, some of the eigenvalues in general could be complex. For illustration of a physical solution, we propose  a `mirror scale difference' through the SSB.  This means the gauge bosons corresponding to $SU(2)^{\prime}_L$ should have the same scale difference with respect to the gauge bosons of the group $SU(2)_R$ as that of the gauge bosons of the gauge group $SU(2)^{\prime}_R$ with respect to the gauge group $SU(2)_L$. The gauge bosons corresponding to $SU(2)^{\prime}_R$ could be at the same or a close scale to $SU(2)_R$.  This means, $v_R = v_{R}^\prime$ and $v_{L}^\prime = 2 v_R$.  It is quite interesting that a physical solution does exists for this symmetry breaking pattern. A more general study of the scalar potential will be provided elsewhere.

With the above assumptions, the equations $\frac{\partial V}{\partial v_L}=\frac{\partial V}{\partial v_R} = 0$ can be used to eliminate $\mu_1$ and $\mu_2$.  After this, we obtain the following constraints on the quartic couplings:
\begin{eqnarray}
\label{quart}
\lambda_1 &=& \dfrac{1}{2} \left[ \lambda_5 + \dfrac{ (\lambda_3 -\lambda_4) (v_L^{\prime 2} -  v_R^2) - (\lambda_9 -\lambda_{10}) (  \omega^2 -  \omega^{\prime2})}{v_L^2 - v_R^{\prime 2}}           \right], \\ \nonumber
\lambda_2 &=& \dfrac{1}{2} \left[ \lambda_6 + \dfrac{ (\lambda_3 -\lambda_4) (v_L^{2} -  v_R^{\prime 2}) +(\lambda_{11} -\lambda_{12}) (  \omega^2 -  \omega^{\prime2})}{v_L^{\prime 2} - v_R^{2}}           \right], \\ \nonumber
\lambda_9 &=&   \dfrac{- (\lambda_{11} v_{R}^2+ \lambda_{12} v_{L}^{\prime 2} )   v_L^2 +  (\lambda_{11} v_{L}^{\prime 2}+ \lambda_{12} v_{R}^{2} )   v_R^{ \prime 2 } - (2 \lambda_{7} v_{L}^{ 2}- \lambda_{8} v_{R}^{\prime 2} )   w^{ 2 } +   (2 \lambda_{7} v_{R}^{\prime 2}- \lambda_{8} v_{L}^{ 2} )   w^{\prime 2 } }{  (v_L^4 - v_R^{\prime 4})}, \\ \nonumber
\lambda_{10} &=&   \dfrac{- (\lambda_{11}  v_{L}^{\prime 2} + \lambda_{12}   v_{R}^2)   v_L^2 +  (\lambda_{11}  v_{R}^{2}  + \lambda_{12}  v_{L}^{\prime 2})   v_R^{ \prime 2 } + (2 \lambda_{7} v_{R}^{\prime 2}- \lambda_{8} v_{L}^{2} )   w^{ 2 } -   (2 \lambda_{7} v_{L}^{2}- \lambda_{8} v_{R}^{\prime 2} )   w^{\prime 2 } }{ (v_L^4 - v_R^{\prime 4})}.
\end{eqnarray}

For the determination of the eigenvalues of Eq.( \ref{sec_der}), we encounter an order five polynomial (since one eigenvalue is always zero) which is  difficult to solve.  For simplifying our calculations, we first expand this polynomial in terms of parameter $\epsilon = v_R / \omega^\prime$ (since $\omega^\prime >> v_R$) and keep only the leading order contribution.  This factorizes our polynomial into three parts, two linear terms and one cubic term.  From linear terms, we obtain the following two scalar masses squared:

\begin{eqnarray}
M_1^2 & =& \dfrac{4}{3} \left(  ( \lambda_3 - \lambda_4) v_L^2 + (\lambda_{11} - \lambda_{12}) (w^2 - w^{\prime 2}) \right) + \mathcal{O} (\epsilon), \\ \nonumber
M_2^2 & =& \dfrac{1}{3} \left(  ( \lambda_3 - \lambda_4) v_L^2 + (\lambda_{11} - \lambda_{12}) (w^2 - w^{\prime 2}) \right) + \mathcal{O} (\epsilon).
\end{eqnarray}
We further expand the cubic part in term of the parameter $\epsilon^\prime = \omega  / \omega^\prime$ (since $\omega^\prime >>  \omega $).  This provides us a quadratic factor and one eigenvalue zero.  Solving quadratic factor, we obtain the physical masses of two other scalars.  These are

\begin{eqnarray}
M_{3,4}^2 &=& \dfrac{1}{2 v_L^2} \left(  x_1 \pm \sqrt{x_2}\right) + \mathcal{O} (\epsilon, \epsilon^{\prime}),
\end{eqnarray}
where
\begin{eqnarray}
x_1 &=& \lambda_5 v_L^4 + 2 \lambda_7 v_L^2 \omega^{\prime 2} + 2 \lambda_7 \omega^{\prime 4} -  \lambda_8 \omega^{\prime 4}  ), \\ \nonumber
x_2 &=& 8 \lambda_7 v_L^2 \left( - \lambda_5  v_L^4 \omega^{\prime 2} + 
  \lambda_8  \omega^{\prime 6}  \right)+ \left(\lambda_5 v_L^4 - \lambda_8 \omega^{\prime 4} + 2 \lambda_7 \omega^{\prime 2} (v_L^2 + \omega^{\prime 2}) \right)^2.
\end{eqnarray}
The condition that all masses squared must be positive and $\omega^\prime >> v_L, v_R, \omega$ implies that
\begin{eqnarray}
\label{minc}
\lambda_{12}  &>& (\lambda_3 v_L^2 - \lambda_4 v_L^2 + \lambda_{11} \omega^2 - \lambda_{11} \omega^{\prime 2})/(w^2 -  \omega^{\prime 2}), \\ \nonumber
\lambda_7 &>& 0, \lambda_8  < \dfrac{\lambda_5 v_L^4}{ \omega^{\prime 4}}.
\end{eqnarray}
The couplings $\lambda_{3}$, $\lambda_{4}$, $\lambda_{5}$, $\lambda_{6}$, $\lambda_{8}$, $\lambda_{11}$ and $\lambda_{12} $ can be positive as well as negative satisfying Eq. (\ref{minc}) appropriately.  For instance, for $v_L = 246$ GeV, $v_R = v_R^\prime = \omega= 1$ TeV and  $v_L^\prime = 2$ TeV, a set of values of quartic couplings which provide a true minimum could be $\lambda_3 = 0.78$, $\lambda_4 = 0.005$,  $\lambda_5 = 0.05$,  $\lambda_{6} = 0.0001$, $\lambda_{7} = 2.0 \times 10^{-12}$, $\lambda_{8} = - 10^{-12}$, $\lambda_{11} =-  0.9$   and $\lambda_{12} =-0.9$.

The scalar potential of the model  does not have any complex coupling.  The gauge symmetry of the model allows us to make the VEVs of the scalar fields real.  Hence, as discussed in Ref. \cite{Barr:1991qx}, the strong $CP$ phase is zero in this model.

Now, we discuss the naturalness of the SM Higgs mass. The one-loop contributions to the mass of the SM Higgs due to fermions is absent, and the contribution begins at three loops.   We calculate the quadratic divergence within the dimensional regularization scheme.  The quadratic divergence is identified with the $D=2$ pole \cite{Veltman:1980mj}.  $h_R$, $S$ and $S^\prime$ denote the Higgs particles corresponding to scalar fields $\varphi_R$, $\chi$ and $\chi^\prime$, respectively.  The quadratic divergent part is given as

\begin{eqnarray}
\label{qd1}
&-&  \int \frac{d^D k_1}{(2 \pi)^D}  \frac{d^D k_2}{(2 \pi)^D}  \frac{d^D k_3}{(2 \pi)^D} Tr\Bigl[ \frac{i (\slashed k_1 + m_f)}{k_1^2-m_f^2} \frac{i}{(k_2^2 - m_{h_R}^2)} \frac{i}{k_3^2-m_S^2}  \\ \nonumber
&& \times  \frac{i  (\slashed p+ \slashed k_1+\slashed k_2+\slashed k_3 + m_f)}{(p+k_1+k_2+k_3)^2-m_f^2} \Bigr] 
  (-i \Gamma_f)  (-i \Gamma_f)
=  \frac{6 i}{(16 \pi^2)^3}  \Lambda ^2 \Gamma_f^2 + \cdots,
\end{eqnarray}

where $\Gamma_f$ denotes the coupling of  fermions running in the loop to the SM Higgs and other scalars.  The noteworthy consequence of the loop suppression is that the quadratic divergent contribution is naturally suppressed.  However, there is no reason that $\Gamma_f$ should be of order $\mathcal{O} (1)$ now.  The mass of the top quark, from Eq.(\ref{mass1}), is given by $m_t = \Gamma_t v_L v_R \omega/2 \sqrt{2} \Lambda^2$.  Hence, even if $\Gamma_t < 1$, the other unknown VEVs could be such that we recover experimental mass of the top quark.  Therefore, the quadratic divergence could be even suppressed further. The analogous contribution to the SM Higgs mass  in the SM, assuming it an effective theory, turns out to be the most  dangerously divergent one.

The one-loop quadratic divergent contribution to the SM Higgs mass which involves scalar doublets and singlets,  is the following:

\begin{eqnarray}
\label{qd2}
\frac{1}{2}  \int \frac{d^D k}{(2 \pi)^D}  \frac{i}{k^2-m_{h_{R},h_{R,L}^\prime, S,S^\prime}^2} (-i \lambda_{3,4, 5,9,10})
= \frac{-i \Lambda ^2}{16 \pi^2} \frac{1}{2} \lambda_{3,4, 5,9,10} + \cdots,
\end{eqnarray}
where $\lambda_{3,4, 5,9,10}$ are couplings of the SM Higgs to the other scalar doublets and singlets.

These contributions depend on the sign and values of the quartic couplings $\lambda_{3,4, 5,9,10}$.   However,  $\lambda_{3,4, 5,}$ can be positive as well as  negative as discussed earlier.  The values of couplings $\lambda_{9,10}$ depend on the values of other couplings as given in Eq.( \ref{quart}).  In principle, they could also be positive as well as  negative.  The contribution to the SM Higgs mass from the scalar doublets and singlets could be  such that they cancel the one-loop quadratic corrections coming from the SM gauge bosons.  Thus, this model 
postpones the fine-tuning of the mass of the SM Higgs up to a scale which is relatively higher than what is obtained assuming an effective SM. In fact, it is known that addition of real scalar singlets to the SM can stabilize the SM Higgs mass up to a sufficiently high scale\cite{singlet-veltman}.  However, such a complex  investigation using the Veltman condition\cite{Veltman:1980mj} is beyond the scope of this paper.

\begin{figure*}[t!]
   \includegraphics[scale=0.5]{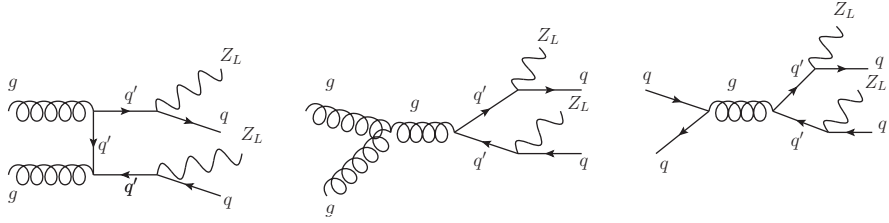}
    \caption{The pair production of the mirror quarks at the LHC and their subsequent decay to the SM $Z_L$ boson and a quark.}
    \label{fig2}
\end{figure*}

The phenomenological signatures of the model will be discussed now.  For this purpose, the charged current Lagrangian can be written as
\begin{eqnarray}
{\cal L}_{CC} &=&-\frac{g_1}{2\sqrt 2} \sum\limits_{F=f, f^\prime} \bar F \gamma^\mu \Bigl[C_{FF}^{W_L}(1-\gamma^5)W_{L\mu}^{-}+C_{FF}^{{W_R^\prime}}(1+\gamma^5){W}_{R\mu}^{\prime-} \Bigr]  F\\ \nonumber
&-& \frac{g_2}{2\sqrt 2} \sum\limits_{F=f, f^\prime}   \bar F \gamma^\mu \Bigl[C_{FF}^{W_R}(1+\gamma^5)W_{R\mu}^{-}+C_{FF}^{{W_L^\prime}}(1-\gamma^5){W}_{L\mu}^{\prime-} \Bigr] F,
\label{eq:cc}
\end{eqnarray}
where the couplings $C_{FF}^{W}$ depend on the charged fermion mixing angles $\theta_L~{\rm and}~\theta_R$. The neutral current Lagrangian is given as 
\begin{eqnarray}
{\cal L}_{NC}  &=& - e Q_f  \sum\limits_{F=f, f^\prime} \bar F \gamma^\mu A_\mu F  \nonumber\\
&-& g_1  \sum\limits_{F=f, f^\prime}  \bar F \gamma^\mu \left[ \left( A_{FF}^{Z_L} \frac{1-\gamma^5}{2}+B_{FF}^{Z_L} \frac{1+\gamma^5}{2} \right)Z_{L \mu}  + \left( A_{FF}^{Z_R^\prime} \frac{1-\gamma^5}{2}+B_{FF}^{Z_R^\prime} \frac{1+\gamma^5}{2} \right)Z_{R \mu}^\prime  \right]  F \nonumber\\
&-& g_2  \sum\limits_{F=f, f^\prime}  \bar F \gamma^\mu \left[ \left( A_{FF}^{Z_R} \frac{1-\gamma^5}{2}+B_{FF}^{Z_R} \frac{1+\gamma^5}{2} \right)Z_{R \mu}  + \left( A_{FF}^{Z_L^\prime} \frac{1-\gamma^5}{2}+B_{FF}^{Z_L^\prime} \frac{1+\gamma^5}{2} \right)Z_{L \mu}^\prime  \right] F,
\label{eq:NC}
\end{eqnarray}
where $e$ is electron charge and $Q_{f,f^\prime}$ is the charge of fermion $f$ and $f^\prime$. The couplings $A_{FF}^Z$ and $B_{FF}^Z$ are functions of  charged fermion mixing angles and gauge mixing angles given in Eqs. (\ref{wth}) and (\ref{fth}).  We observe from charged and neutral current Lagrangians that the mirror quarks can decay into a $W_L$ or $Z_L$ boson in association with a SM quark.  Moreover, the mirror quarks can decay into the SM Higgs and a SM quark.    For illustration, we show the pair production of the mirror quarks in Fig.\ref{fig2} at the LHC via gluon-gluon and quark-antiquark initial states.  

In addition to this, flavor changing neutral meson mixings  $K-\bar{K}$, $B-\bar{B}$ and $D-\bar{D}$ will further put constraints on new gauge bosons.  The masses of the gauge bosons corresponding to the gauge groups $SU(2)_R$ in the minimal left-right symmetric model (MLRSM) \cite{Mohapatra:1979ia} are highly constrained by the mixing of $W_L$ and $W_R$ bosons.  The masses of these gauge bosons are excluded up to approximately 3 TeV \cite{Bertolini:2014sua} in this model.  Since, there is no mixing between $W_L$ and $W_R$ bosons in the proposed model, this limit is not applicable.  A detailed phenomenological investigation is the subject of a future study.

Now we comment on the importance of the singlet scalar fields.  It should be noted that singlet scalar fields are not arbitrarily introduced in this model.   They have a rather important purpose to provide masses to fermions.    Furthermore, the singlet scalar fields could be a viable cold dark matter candidate as observed in some models \cite{Cdark}.  Finally, they could make the electroweak phase transition a strong first-order transition \cite{ewpt}.

The ultraviolet completion of the model could come from a larger underlying theory.  Since, there are many VEVs in this model, one of the possibilities is that this model could be a part of a multiverse theory with many ground states which is motivated by the fine-tuning of the cosmological constant\cite{Donoghue:2016tjk}.  This is encouraged by the fact that  there are two real massless scalars in the model coming from two real singlet scalar fields.  The quantum and/or thermal fluctuations in the early universe would randomize the initial values of these fields leading to regions of different initial values due to inflation.  We comment why we have chosen real scalar singlets massless in the scalar potential now.  The multiverse theories require a continuous variation of the parameters across the universe\cite{Donoghue:2016tjk}.  Only a field can have a spatial or temporal variation. For this purpose, that field must be light\cite{Donoghue:2016tjk}. This is the case for real scalar singlets in this model.

This model restores parity in a nonminimal way.  We observe that parity is maximally violated in the SM.  Now, maximal violation of parity could be a consequence of a minimal or maximal parity restoring theory.  Our approach in this work is that the maximal parity violation leads to a maximal parity restoring theory. We note that MLRSM has a  VEV which must be zero or vanishing to reproduce neutrino masses.  This is similar to unnaturally small Yukawa couplings of neutrinos in the SM extended by three right-handed singlet neutrinos.  The other mirror models seem to have a huge scale disparity in the gauge sector\cite{Foot:1991bp,Silagadze:1995tr,Foot:1995pa,Berezhiani:1995yi,Gu:2012in,Chakdar:2013tca}.  Furthermore, they do not have any explanation for the smallness of neutrino masses. Moreover, the above two classes of models  do not have any mechanism to make the mass of the SM Higgs natural. Therefore, if one discards the prejudice of minimality,  the model presented in this work is a natural parity restoring extension of the SM.

\section*{Acknowledgements}
It is my immense pleasure to thank Antonio Pich for very useful discussion during this work.  This work has been supported by the Spanish Government and ERDF funds from the EU Commission
[Grants No. FPA2011-23778, FPA2014-53631-C2-1-P and No. CSD2007-00042 (Consolider Project CPAN)].



\end{document}